\documentclass[12pt,a4paper,dvips]{article}
\usepackage{cite,mcite}
\usepackage{graphicx}
\newlength{\capindent}
\newlength{\capwidth}
\newlength{\figwidth}
\newcommand{\icaption}[2][!*!,!]{\hspace*{\capindent}%
  \begin{minipage}{\capwidth}
    \ifthenelse{\equal{#1}{!*!,!}}%
      {\caption{#2}}%
      {\caption[#1]{#2}}
  \end{minipage}}
\def\ee{\ensuremath{\mathrm{e^+ e^-}}}%
\newcommand{\EE}{\ensuremath{\mathrm{e}^+ \mathrm{e}^-}}
\newcommand{\FF}{\ensuremath{f\bar f}}
\newcommand{\EEFF}{\ensuremath{\EE\rightarrow\FF}}
\def\mm{\ensuremath{\mathrm{\mu^+ \mu^-}}}%
\def\tautau{\ensuremath{\mathrm{\tau^+ \tau^-}}}%

\newcommand{\TeV}{\ensuremath{\mathrm{Te\kern -0.12em V}}}
\def\GeV{\ifmmode {\mathrm{\ Ge\kern -0.1em V}}\else
  \textrm{Ge\kern -0.1em V}\fi}%

\newcommand {\Be}{\begin{equation}}
\newcommand {\Ee}{\end{equation}}

\newcommand {\eqref}[1]{equation~(\ref{#1})}

\renewcommand{\thefootnote}{\fnsymbol{footnote}}

\begin{document}

\begin{titlepage}
\begin{picture}(100,-100)(10,20)
\put (245.,112.){\bf June 2, 2025}
\end{picture}

\vspace*{2.0cm}

\begin{center} {\Large \bf
       Muon (and Lepton) Anomalous Magnetic Moments\\
\vspace*{0.15cm}
       and Limits on Their Radii}

\vspace*{2.0cm}
  {\Large
  Dimitri Bourilkov\footnote{\tt e-mail: dimi@ufl.edu}
  }

\vspace*{1.0cm}
  University of Florida \\
  Gainesville, Florida, 32611, USA
\vspace*{3.0cm}
\end{center}

%
%
\begin{abstract}
The limits on lepton radii and contact interaction scales are used to
derive limits on the deviations of lepton anomalous magnetic dipole
moments a=(g-2)/2 from the Standard Model predictions. In the case of
quadratic dependence of the deviations on the lepton mass, the limits
for electrons are better compared to low energy measurements, for
muons somewhat weaker than the latest precision experiments, and for
taus by far the best.
\end{abstract}


\end{titlepage}

\renewcommand{\thefootnote}{\arabic{footnote}}
\setcounter{footnote}{0}
%
%

Without a doubt the Standard Model (SM) is very successful in
describing the precision measurements performed at the
highest energy accelerators. The SM assumes that fundamental
particles, like the leptons, are point-like.

At the energy frontier, the lepton-pair measurements at LEP
at energies above the Z resonance mass provide very sensitive
probes of the point-like structure of the
electron~\cite{Bourilkov:2000ap,Bourilkov:2001pe},
the muon, and the tau~\cite{Acciarri_2000}.

At the precision frontier, the Fermilab Muon g-2
experiment~\cite{PhysRevD.110.032009,collaboration2025measurementpositivemuonanomalous}
has generated excitement when comparing their measurements
to the best available theory predictions. If the anomalous
magnetic moments deviate from theory, this may be a sign
for lepton substructure and finite lepton radii, or for new
interactions. In this study we analyze the interplay between
lepton radii, contact interactions and anomalous magnetic
moments. 

%
%

The high precision measurements of the magnetic dipole moment
$\rm g_e$ of the electron have been used to put stringent
limits on the electron radius $\rm r_e$~\cite{Brodsky,Kopp:1994qv}.
If non-standard contributions to $\rm g_e$ scale linearly
with the electron mass, the most precise bound~\cite{Fan:2022eto}
is below $\rm 10^{-24}\ m$.
As a natural consequence of chiral symmetry,
these corrections are expected to scale quadratically with
the lepton mass~\cite{Brodsky}. In this case the bound
is reduced to $\rm r_e < 3.2\cdot 10^{-19}\ m$.

The L3 collaboration has analyzed the fermion-pair measurements
above the Z pole for deviations from the point-like
structure~\cite{Acciarri_2000}.
A combined analysis of the electron data is also
available~\cite{Bourilkov:2000ap,Bourilkov:2001pe}.
The SM cross sections for the reactions {\EEFF}
are modified as follows:
\begin{equation}
 \frac{d \sigma}{d q^2} = \left(\frac{d \sigma}{d q^2}\right)_{SM} F_e^2(q^2) F_f^2(q^2),
\end{equation}
where $ q^2$ is the Mandelstam variable $s$ or $t$ for
$s$-- or $t$--channel exchange, the latter only important for Bhabha
scattering, and the form factors of the initial and final state fermions
are denoted as $ F_e$ and $ F_f$, respectively.
They are parameterized by a Dirac form factor:
\begin{equation}
 F(q^2) = 1 + \frac{1}{6} q^2 r^2,
\end{equation}
where $ r $ is the radius of the fermions.

The upper limits on the fermion radii 
obtained from the LEP data are shown in Table~\ref{tab:fradii1}.
They are derived with the assumption $F_e = F_f$.
For the {\mm} and  {\tautau}
final states the limits given in Table~\ref{tab:fradii1}
will increase by a factor of $\sqrt{2}$ under the
most conservative assumption that the electron is
point-like ($F_e \equiv 1$).


%
\begin{table}[b]
 \renewcommand{\arraystretch}{1.2}
  \begin{center}
    \begin{tabular}{|c|c|c|}
\hline
~~Channel~~&~~~$r$  [m]~~&~~~$\delta a_l$~~\\
\hline 
\ee\       & $\rm 2.8\cdot 10^{-19}$ & $\rm 5.3\cdot 10^{-13}$ \\
\mm\       & $\rm 2.4\cdot 10^{-19}$ & $\rm 1.7\cdot 10^{-8}$  \\
\tautau\   & $\rm 4.0\cdot 10^{-19}$ & $\rm 1.3\cdot 10^{-5}$  \\
\hline
    \end{tabular}
  \end{center}
  \caption{
    Upper limits on the fermion radii at 95~\% confidence level
    from form factors
    for electrons~\protect\cite{Bourilkov:2000ap},
    muons and taus~\protect\cite{Acciarri_2000},
    and corresponding values for $\delta a_l$.
    }
  \label{tab:fradii1}
\end{table}

From the combined analysis\cite{Bourilkov:2000ap}
of the LEP collaborations data on Bhabha scattering the
best upper limit on the electron radius at 95\% confidence
level was derived:
\Be
\rm r_e < 2.8\cdot 10^{-19}\ m.
\Ee

This limit is below the limit
derived from $\rm (g-2)_e$ measurements in the case where
the deviations from the SM of the magnetic dipole moment of
the electron depend quadratically on its mass.

To connect the bounds on lepton radii to deviations for
the anomalous magnetic moment we will use a quadratic
dependence on the masses:
\Be
 \delta a_l = m_l^2r_l^2
\Ee
where $\delta a_l$ is the deviation of the anomalous magnetic
dipole measurement from the SM prediction for lepton $l$,
$m_l$ is the lepton mass in \GeV, and $r_l$ is the lepton
radius in units 1/\GeV. To help the reader we convert
most distances to meters.

%
%

In searches for new interactions beyond the SM, a standard framework
is the most general combination of helicity conserving dimension-6
operators~\cite{PeskinCI}. In this scheme, we have as parameters
a coupling strength, $g$, and an energy scale,
$\Lambda$, which can be viewed as the scale of compositeness. 
At energies well below $\Lambda$, this results in an effective Lagrangian
leading to four-fermion contact interactions.

As the LEP experiments have observed no statistically significant
deviations from the SM, they have derived one-sided
lower limits on the scale $\rm \Lambda$ of contact interactions at
95\% confidence level~\cite{ALEPH:2013dgf}
under the assumption of a strong
coupling $\frac{g^2}{\rm{4\pi}}=1$.
The best limits are summarized in Table~\ref{tab:fradii2}.

\begin{table}[h]
 \renewcommand{\arraystretch}{1.2}
  \begin{center}
    \begin{tabular}{|c||c|c||c|c|}
\hline
~~Channel~~&~~~~$~~~\Lambda_{VV}^+$~~&~~~$r$~~&~~~$\lambda_{VV}^+$~~&~~~$r$~~\\
                &~~~[TeV]            &~~~[m]  &~~~[TeV]             &~~~[m]  \\            
\hline 
\ee\       & 21.3   & $\rm 9.2\cdot 10^{-21}$ & 1.9    & $\rm 1.0\cdot 10^{-19}$  \\
\mm\       & 18.9   & $\rm 1.0\cdot 10^{-20}$ & 1.7    & $\rm 1.2\cdot 10^{-19}$  \\
\tautau\   & 15.8   & $\rm 1.2\cdot 10^{-20}$ & 1.4    & $\rm 1.4\cdot 10^{-19}$  \\
\hline
    \end{tabular}
  \end{center}
  \caption{
    Lower limits at 95~\% confidence level for
    electrons~\protect\cite{Bourilkov:2001pe},
    muons and taus~\protect\cite{ALEPH:2013dgf}
    on the contact interaction scales $\Lambda$ for the VV
    model with positive interference, and for the corresponding
    ``QED rescaled'' scales $\lambda$. The scales are converted
    to upper limits on the fermion radii.}
  \label{tab:fradii2}
\end{table}

%
%

How to compare the limits on lepton radii to the limits on the contact
interaction scale? They are different parts of the same puzzle.  For
example, the limit on the electron size
\mbox{$\rm r_e < 2.8\cdot  10^{-19}\ m$}
translates to a mass scale \mbox{$\rm M > 0.7\ TeV$}.

In contrast, for contact interactions very stringent limits
$\sim$~15-20~TeV were derived. By convention, a strong coupling
$\frac{g^2}{\rm{4\pi}}=1$ for the novel interactions is postulated, which
may be too optimistic. In a more conservative approach, if we assume
a coupling of electromagnetic strength, the limits are reduced
substantially:
\newpage
\Be
 \rm \lambda = \sqrt{\alpha_{QED}}\cdot \Lambda = 0.0884 \cdot \Lambda
\Ee
where the value of
$\rm \alpha_{QED}$ at the Z resonance is used and small effects, at
the percent level, of the running of $\rm \alpha_{QED}$ to higher
energies are ignored. $\lambda$ is the ``QED modified'', lower, scale.
A careful study shows that the VV contact interaction model with positive
interference modifies the differential cross section in a similar way
as a form factor, so we will use these results for comparisons.  The
results for $\Lambda$ and $\lambda$ are compiled in Table~\ref{tab:fradii2}.
As can be seen, in the case of $\lambda$ the bounds on the lepton radii
are close, and lower, than the form factor bounds.

The limits on $\delta a_l$ coming from form factors are summarized in
Table~\ref{tab:fradii1}, and from contact interactions for $\lambda$
in Table~\ref{tab:fradii3}. Under the quadratic dependence assumption,
our electron limits are better than the (g-2) measurements.

The situation is reversed for muons, where the latest
comparison~\cite{PhysRevD.110.032009,aliberti2025anomalousmagneticmomentmuon}
gives
\Be
 \delta a_\mu = 2.6\pm6.6\cdot 10^{-10}
\Ee
compared to our limits
\Be
 \delta a_\mu = 1.7\cdot 10^{-8} \ \ \rm{for\ form\ factor}
\Ee
\Be
 \delta a_\mu = 4.1\cdot 10^{-9} \ \ \rm{for\  contact\ interaction}.
\Ee
If a novel interaction happens to exceed electromagnetic strength,
the limit can become more competitive.

The situation is reversed for tau leptons: as experimental
measurements~\cite{ParticleDataGroup:2024cfk}
of $\delta a_\tau$ are very difficult, our limits by far exceed
direct approaches.

\begin{table}[h]
 \renewcommand{\arraystretch}{1.2}
  \begin{center}
    \begin{tabular}{|c|c|c|}
\hline
~~Channel~~&~~~$r$  [m]~~&~~~$\delta a_l$~~\\
\hline 
\ee\       & $\rm 1.0\cdot 10^{-19}$ & $\rm 6.7\cdot 10^{-14}$ \\
\mm\       & $\rm 1.2\cdot 10^{-19}$ & $\rm 4.1\cdot 10^{-9}$  \\
\tautau\   & $\rm 1.4\cdot 10^{-19}$ & $\rm 1.6\cdot 10^{-6}$  \\
\hline
    \end{tabular}
  \end{center}
  \caption{
    Upper limits on the fermion radii at 95~\% confidence level
    from contact interactions with scale $\lambda$
    for electrons, muons and taus, and corresponding values for
    $\delta a_l$}
  \label{tab:fradii3}
\end{table}

To conclude: we used the limits on lepton radii and contact
interaction scales to derive limits on the deviations of lepton
anomalous magnetic dipole moments (g-2)/2 from the Standard Model
predictions. If the deviations exhibit a quadratic dependence on
the lepton mass, the limits for electrons are better compared to low
energy measurements, for muons somewhat weaker than the latest
precision experiments, and for taus by far the best.

%
%
\section*{Acknowledgments}
The author is grateful to the University of Florida High Energy Group
for supporting this type of work.

%
%

\newpage
\bibliographystyle{l3stylem}
\bibliography{g-2-2025}

\end{document}